\documentclass[12pt]{article}
\usepackage{amsmath}
\usepackage{graphicx}
\usepackage{amsfonts}
\usepackage{amssymb}
\usepackage{epsfig}

\newcommand{\kor}{\kappa_{ \rm orb}}
\newcommand{\ks}{\kappa_{6}}
\newcommand{\kt}{\kappa_{10}}
\newcommand{\ap}{{\alpha^{\prime}}}
\newcommand{\zb}{\bar{z}}
\newcommand{\esp}{{\rm e}}
\newcommand{\tr}{{\rm Tr}}
\newcommand{\no}{\nonumber}
\newcommand{\nod}{\nonumber \\}
\newcommand{\nodu}{\nonumber \right. \\ & & \left.}

\newcommand{\inn}{\!\cdot\!}
\newcommand{\G}{\Gamma}
\newcommand{\pol}{\varepsilon}
\newcommand{\ten}[2]{{\rm #1}_{(#2)}}
\newcommand{\tenn}[1]{{\rm #1}}
\newcommand{\start}{^{*_{10}}}
\newcommand{\stars}{^{*_{6}}}
\newcommand{\bnp}[1]{Nucl. Phys. {\bf #1}}

\begin{document}
\begin{titlepage}
\rightline{DFTT-48/2000}
\rightline{DSF-39/2000}
\rightline{hep-th/0012193}
\vskip1.2cm
\centerline{\Large\bf World Volume Action for Fractional Branes $^\ast$}
\vskip1.2cm
\centerline{\bf Paolo Merlatti}
\centerline{\sl Dipartimento di Fisica Teorica, Universit\`a di Torino}
\centerline{\sl and I.N.F.N., sezione di Torino}
\centerline{\sl via P. Giuria 1, I-10125 Torino, Italy}
\centerline{\sl Paolo.Merlatti@to.infn.it}

\vskip1cm
\centerline{\bf Gianluca Sabella}
\centerline{\sl Dipartimento di Scienze Fisiche, Universit\`a di Napoli
``Federico II"}
\centerline{\sl and I.N.F.N., sezione di Napoli}
\centerline{\sl Complesso
Universitario di Monte S. Angelo, Via Cintia, I-80126 Napoli, Italy}
\centerline{\sl
Gianluca.Sabella@na.infn.it}
\vskip2cm
\begin{abstract}
We study the world volume action of fractional D$p$-branes of type IIA string theory
compactified on the orbifold ${T^4}/{Z_2}$. The geometric relation between these branes
and wrapped branes is investigated using conformal techniques. In particular we examine
in detail various scattering amplitudes and find that the leading low-energy interactions
are consistent with the boundary action derived geometrically.
\end{abstract}

\vfill \hrule width 6cm \vskip2mm
{\footnotesize $^\ast$ Work partially supported by the EC
RTN programme HPRN-CT-2000-00131 in which G.S. is associated with Frascati-LNF.}

\end{titlepage}

\newpage

\tableofcontents


\section{Introduction}

Maldacena's duality \cite{9711200} provides a remarkable relation between string theory
and supersymmetric and conformal quantum field theory. Recently a lot of effort has been
devoted to extend this duality and find new correspondences between string theories and
non conformal and less supersymmetric gauge theories. These attempts have been developed
in several directions, one of which is the study of fractional D-branes \cite{9712230} on
conifold \cite{0007191} and orbifold singularities \cite{0011077,0012035}. In this case
the dual gauge theory corresponding to a stack of M such fractional branes is non
conformal and supersymmetry is partially broken with respect to the original case
considered by Maldacena.

It is well known that the fractional D$p$-branes can be viewed as ordinary
D($p+2$)-branes wrapped on an integer basis of vanishing two-cycles \cite{9712230}. In
this paper we want to test this correspondence by explicit calculations of string
scattering amplitudes. In particular we focus on the fractional D$p$-branes in type IIA
string theory compactified on $T^4/\mathbf{Z_{2}}$, where $\mathbf{Z_2}$ is generated by
the parity operator on the four compact spatial coordinates. More specifically, we want to
check that the boundary action for fractional D$p$-branes, obtained geometrically from
the one of D($p+2$) branes suitably wrapped, is in agreement with the CFT predictions. To
do this we first determine the supergravity fields which couple to the fractional D-brane
by means of the boundary state formalism (for a review see \cite{anto}). This simply
amounts to obtain the terms of the world-volume action of the fractional D-brane which
are linear in the bulk fields. Then, we go on and compute the quadratic terms of the
fractional branes world-volume action using the same methods that have been discussed
for ordinary D-branes in \cite{9808074,9812149,0004198,9603194,9611214,9809100,9901085}.\\
These calculations confirm that the world-volume action obtained geometrically is
consistent with the explicit conformal calculations.

This paper is organized as follows: in section 2 we review the geometric approach to
fractional D-branes and derive accordingly their world-volume action; in section 3 we
calculate the linear and quadratic scattering amplitudes between the massless fields that
couple to fractional D-branes. The three appendices are devoted respectively to fix our
notation, determine the normalization for the boundary state and to list the vertices
used in various calculations.


\section{The geometric approach to the fractional Dp brane action}

The action for type IIA supergravity in ten dimensions  can be written (in the string
frame) as:
\begin{eqnarray}
S_{\rm IIA} &=&
	\frac{1}{2 \kt^2}
	\Bigg\{
		\int d^{10} x ~
		\sqrt{\rm -det G}~\esp^{-2\Phi} ~ {\rm R(G)}
		+ \int \esp^{-2\Phi}
		\Big[ ~
			4 d \Phi \wedge ~ \start d \Phi ~
			-\frac{1}{2} \ten{H}{3} \wedge ~ \start \ten{H}{3}\Big]
			\nod
&&
			+ \frac{1}{2}\Big[ ~ \ten{F}{2} \wedge ~ \start \ten{F}{2} ~
			+ ~ \ten{\widetilde{F}}{4} \wedge ~ \start \ten{\widetilde{F}}{4} ~
			-\ten{B}{2}\wedge\ten{F}{4}\wedge\ten{F}{4}~
		\Big]
	\Bigg\}
\label{bulk}
\end{eqnarray}
where
\begin{eqnarray}
\label{form2}
 \ten{H}{3}=d \ten{B}{2} \ , \hskip1cm
 \ten{F}{2}=d \ten{C}{1} \ , \hskip1cm
 \ten{F}{4}=d \ten{C}{3}
\end{eqnarray}
are respectively the field strengths corresponding to the NS-NS 2-form potential, to the
1-form and 3-form potentials of the R-R sector; moreover:
\begin{eqnarray}
	\ten{\widetilde{F}}{4} &=&
		\ten{F}{4} - \ten{C}{1} \wedge \ten{H}{3} \ ,
   \label{form3}
\end{eqnarray}
and
\begin{eqnarray}
   \no\kappa_{10} &=& 8\,\pi^{7/2}\,g_s\,\ap^{2}
\end{eqnarray}
where $g_s$ is the string coupling constant.

The usual BPS D$p$-branes with $p$ even are solutions of the classical field equations
that follow from the action (\ref{bulk}), which are charged under the R-R $(p+1)$-form
potentials and preserve sixteen supercharges. On the other hand, the interaction between
the massless fields and a D$p$ brane is described by the boundary action (also in the
string frame):
\begin{eqnarray}
 S  &=&
	-\frac{T_{p}}{\kt}\Bigg \{
	\int d^{p+1}\xi ~ \esp^{-\tenn{\Phi}}
	\sqrt{-{\rm det [
		\tenn{\tilde{G}}+\tenn{\tilde{B}}
		+2\pi \ap \tenn{F}]}} \nod
&&
   \hskip1cm - \int ~ \Big[ \esp^{\tenn{B}+2\pi\ap\tenn{F}}\wedge\sum_n \ten{C}{n+1}\Big]_{p+1}
   \Bigg \} \ .
\label{BI}
\end{eqnarray}
where $T_{p}/{\kt}$ is the D-brane tension, $T_p=\sqrt{\pi}(2\pi\sqrt{\ap})^{3-p}$ and the
tilde denotes the pullback.

Let us now consider type IIA supergravity compactified on the orbifold
\begin{eqnarray}
 \label{bg}
 \mathbb{R}^{1,5}\times T^4/\mathbf{Z_{2}}
\end{eqnarray}
where ${\bf Z_2}$ is the reflection parity that changes the sign to the four coordinates
of the torus $T^4$, which we take to be $x^6$, $x^7$, $x^8$ and $x^9$. The bulk action
for this theory is still given by (\ref{bulk}) suitably compactified \cite{0012035}, but
with $\kt$ replaced by $\kor=\sqrt{2}\kt$ . \\ The orbifold (\ref{bg}) is a singular
limit of a smooth K3 surface. In this case, besides the usual D$p$-branes (bulk branes)
which can freely move in the compact directions, there are \textit{fractional} D$p$-branes
\cite{9712230} which are instead constrained to stay at one of the orbifold fixed points.
The fractional D$p$-branes can be viewed as ordinary D($p+2$)-branes wrapped on an
integer basis of cycles $c_{I} \in \mathrm{H_{2}}({\rm K3},\mathbb{Z})$, with a non
trivial $B$-flux, in the collapsing limit in which these cycles shrink \cite{9611137}. We
study this theory at one fixed point, with fractional branes that extend only in the
space-time directions. In this context it is enough to consider only one cycle for each
fixed point. By thinking of the resolved space in the neighborhood of one fixed point,
the geometrical counterparts of the twisted gauge fields \cite{9712230} are the
Kaluza-Klein $p+1$ forms $A_{p+1}$ that come from the harmonic decomposition
\begin{eqnarray}
 \label{KK}
 \ten{C}{p+3} &=&
	C_{p+3}+\sqrt{V_4} ~ A_{p+1} \wedge \hat{\omega}
\end{eqnarray}
where $C_{p+3}$ and $A_{p+1}$ are six-dimensional forms (from now on we will use italic
style for the six dimensional fields), while $\hat{\omega}$ is the anti-self-dual (1,1)
form, which is Poincar\`e dual to the relevant cycle $c$ and it's normalized as follows:
\begin{eqnarray}
 \int_{{\rm K3}}\hat{\omega}\wedge^{\star_4}\hat{\omega} \ = \ 1 \hskip1cm
 \int_{{\rm c}}\hat{\omega} \ = \  \sqrt{2}
\end{eqnarray}
The factor of normalization in front of $A_{p+1}$ ($V_{4}$ is the volume of the torus
$T^{4}$) in (\ref{KK}) is due to the fact that generally the compactification works
differently for untwisted and twisted fields, but we want to write an homogeneous
six-dimensional effective action. Consider, for example, the kinetic term in the ten
dimensional Lagrangian (\ref{bulk}) for the RR 3-form:
\begin{eqnarray}
 \frac{1}{2 \kor^2}
 \int ~ \frac{1}{2} ~
 d \ten{C}{3} \wedge \start d \ten{C}{3} \ .
 \no
\end{eqnarray}
If we decompose as in (\ref{KK})
\begin{eqnarray}
 \ten{C}{3} &=&
	C_{3}+\sqrt{V_4} A_{1} \wedge \hat{\omega}
 \no
\end{eqnarray}
we get a standard term in the six-dimensional effective Lagrangian
\begin{eqnarray}
 \frac{1}{2 \ks^2}
 \int \frac{1}{2} ~ \Big[
		 ~  d C_{(3)} \wedge \stars (d C_{(3)}) ~
			+ d A_{(1)} \wedge \stars (d A_{(1)})
	  \Big]
\end{eqnarray}
where $\ks=\kor/\sqrt{V_4}$.

Now consider a configuration in which the ten-dimensional two-form $\tenn{B}$ takes a background
value different from zero along the cycle $c$:
\begin{eqnarray}
\label{b}
 \ten{B}{2} &=&
	B_{(2)} + \sqrt{V_4} ~ b \ \hat{\omega}
\end{eqnarray}
and wrap a D($p+2$)-brane on $c$. According to what we have said, the resulting
configuration will be a fractional D$p$-brane and its boundary action can be obtained from
(\ref{BI}) using the wrapping formulas  (\ref{KK}) and (\ref{b}). In the Einstein frame
this action reads:
\begin{eqnarray}
 \label{SBIF}
 S_{{\rm boundary}} &=&
	-\frac{\hat{T}_{p}}{\sqrt{2} \ks}
		\left\{
			\int d^{p+1}x \ \esp^{-\frac{1-p}{2}\phi + \frac{1}{2} \sum_i \eta_i}
			\sqrt{-{\rm det}\,
				\left[
				\tilde{G} + \esp^{-\phi} \tilde{B} + 2 \pi \ap ~\esp^{-\phi}F
				\right]}
			\nodu
			\hskip1.5cm
			- \int ~ \Big[ \esp^{B+2\pi\ap F}\wedge \sum_n C_{n+1}\Big]_{p+1}
		\right\}
	\nod
&&      - \frac{N_{T,p}}{\ks}
		\left\{
			\int d^{p+1} x \
			\esp^{-\frac{1-p}{2}\phi +\frac{1}{2} \sum_i \eta_i} \
			\sqrt{-{\rm det}\,
				\left[
				\tilde{G} + \esp^{-\phi} \tilde{B} + 2 \pi \ap ~\esp^{-\phi} F
				\right]} \
			\tilde{b}
			\nodu
			\hskip1.5cm
			- \int ~ \Big[\esp^{B+2\pi\ap F} \wedge\sum_n \left( C_{n+1} \tilde{b} + A_{n+1}
			\right)\Big]_{p+1}
		\right\}
\end{eqnarray}
where:
\begin{eqnarray}
 \hat{T}_p ~ = ~ \frac{T_{p}}{\sqrt{V_4}}
 \hskip1cm
 N_{T,p} ~ = ~ \frac{\sqrt{2}~ T_{p}}{4\pi^2\ap} \hskip.5cm.
\end{eqnarray}
The six dimensional dilaton field $\phi$ and the scalar fields $\eta_i$ are defined as
follows
\begin{eqnarray}
 \phi &=& \Phi-\frac{1}{4}\ln\Big( \prod_i G_{ii} \Big)
 \hskip1cm
 G_{ii}=\esp^{2\eta_i}
 \hskip.5cm
 \mbox{with } i=6,7,8,9
\end{eqnarray}
while the twisted field $\tilde{b}$ represents the fluctuation part of $b$ around the
background value characteristic of the $\mathbf{Z}_2$ orbifold \cite{9611137,0011060},
which in our notation is
\begin{eqnarray}
 b &=&
 \frac{1}{2}\frac{(2\pi\sqrt{\ap})^2}{\sqrt{V_4}}~ + ~\tilde{b} \ .
\end{eqnarray}
We want to stress that what is really characteristic of this background is the flux of
the field B along the cycle $c$. This does not depend on the normalization of the (1,1)
form $\hat{\omega}$ and it has to take the value $\frac{\sqrt{2}}{2}(2\pi\sqrt{\ap})^2$.

In this paper we want to test the validity of the geometric approach just described by
explicitly computing string scattering amplitudes. To do this, we need the string mode
fluctuations for canonically normalized fields. In our case they take the form (see also \cite{0012035}):
\begin{eqnarray}
\label{edefcanonic}
 g_{\mu\nu} &=& \eta_{\mu\nu}+2k_6 ~ h_{\mu\nu}\no\\
 \phi &=& \ks ~ \varphi\no\\
 B_{\mu\nu} &=& \sqrt{2}\ks ~ b_{\mu\nu}\no\\
 C_{(n)} &=& \sqrt{2}\ks ~ c_{n}\\
 A_{(n)} &=& \sqrt{2}\ks ~ a_{n}\no\\
 \tilde{b} &=& \sqrt{2} \ks ~ \beta\no \\
 \eta_i &=& \ks ~ \hat{\eta}_i\no
\end{eqnarray}

In the next paragraphs we examine all the linear and quadratic couplings of the closed strings
to the D-brane and compare them with the results we can predict from the action
(\ref{SBIF}), finding complete agreement.


\section{String amplitudes for fractional branes}

On general grounds the properties of D-branes can be discussed either by studying their interaction
with open strings \cite{9604065} or by introducing the boundary state (for a review see
\cite{anto}).

The boundary state is a BRST invariant state written in terms of the closed string
oscillators. It encodes the couplings of the D-branes with all states of the closed
string spectrum and inserts a boundary on the closed string world sheet enforcing on it
the appropriate boundary conditions. In an orbifold background like (\ref{bg}) it has
four different components which correspond to the (usual) NS-NS and R-R untwisted sectors
and to the NS-NS and R-R twisted sectors \cite{0011060,9906242,9910109,9912157}, (for a more
detailed analysis see appendix \ref{aboundary}). Generally it provides two essential
information: the couplings with the massless closed string states (from which we read the
mass and the charges) and their
long distance behaviour (from which we study the space time geometry) \cite{9707068}.

We are going to analyze the first aspect in the following subsection.

\subsection{Linear coupling}

The generator of the interaction amplitudes of any closed string state with a fractional
D$p$-brane is the following boundary state:
\begin{eqnarray}
\label{eboundary}
 |{\rm D}p\rangle &=& N_{U,p} ~
	\Big( |B\rangle_{NS,U} \pm |B\rangle_{R,U} \Big)
	\pm N_{T,p} ~
	\Big( |B\rangle_{NS,T} \pm |B\rangle_{R,T} \Big)
\end{eqnarray}
where (see appendix \ref{aboundary}):
\begin{eqnarray}
N_{U,p} &=& \frac{1}{\sqrt{2}}\frac{\hat{T}_p}{2} \\
N_{T,p} &=& \frac{\sqrt{2}~ T_{p}}{4\pi^2\ap} \no
\end{eqnarray}

The various amplitudes can be computed by
simply saturating $|B\rangle$ with the properly normalized closed string states. For
example the twisted NS-NS scalar $\beta$, which is related to the fluctuations of the $b$
field of (\ref{b}), is represented by the following vertex operator:
\begin{eqnarray}
  V_{\beta}^{(-1,-1)} (z,\zb) = \frac{1}{\sqrt{2}}~\beta
		~ C_{ab}S^a(z)\bar{S}^b(\zb) ~
		\esp^{i k\cdot X(z,\zb)} ~
		\esp^{-\phi(z)}\esp^{-\tilde{\phi}(\zb)}~
		\Lambda(z) \tilde{\Lambda}(\zb)
\end{eqnarray}
We have written it in the (-1,-1) superghost picture, since this is the appropriate
picture to use with the boundary state \cite{9707068}. Then the coupling between $\beta$
and a fractional D$p$-brane will be given by:

\begin{eqnarray}
\label{elinbeta}
 \langle ~V_{\beta} ~|~ {\rm D}p~\rangle &=& -\sqrt{2}~ N_{T,p} ~ V_{p+1} ~ \beta
\end{eqnarray}
We see that this coupling is in perfect agreement with what we get from the action
(\ref{SBIF}). We can now proceed analogously for all other massless states that can
couple to the D-brane. Here we simply list the results (see appendix \ref{avertices} for
the explicit expressions of the vertices), since the explicit calculations are identical
to those already discussed in the literature (see e.g. \cite{9707068}):
\begin{eqnarray}
 &
 \langle ~V_{h} ~|~ {\rm D}p~\rangle & = \ - \frac{\Hat{T}_p}{\sqrt{2}}~  V_{p+1} ~ \tr (h\inn L)
 \nod
 &
 \langle ~V_{\hat{\eta}_i} ~|~ {\rm D}p~\rangle  & = \ - \frac{\hat{T}_p}{2\sqrt{2}}~  V_{p+1} ~
 \hat{\eta}_i \nod
 & \langle ~V_{a_{p+1}} ~|~ {\rm D}p~\rangle & = \ \sqrt{2}N_{T,p}~ V_{p+1} ~a_{p+1} \label{res}\\
& \langle ~V_{c_{p+1}} ~|~ {\rm D}p~\rangle & = \ \hat{T_p}~ V_{p+1} ~c_{p+1}\no
\end{eqnarray}
where the fields $h_{\mu\nu}$, $\hat{\eta}_i$, $a_{p+1}$ and $c_{p+1}$ have been defined
in (\ref{edefcanonic}), and the longitudinal matrix $L$ is given by
\begin{equation}
\label{long}
L = \frac{1}{2} \left( 1 + S \right)
\end{equation}
where $S$ is
\begin{eqnarray}
\label{esse}
 S^{\mu}_{\ \nu} &=& \left(
		\begin{array}{cc}
		{\bf 1}_{p+1} & 0 \\
		0 & -{\bf 1}_{5-p}
		\end{array}
	   \right)
\end{eqnarray}
and encodes the Neumann and Dirichlet boundary conditions. For later convenience we also
introduce the matrix N
\begin{equation}
\label{enne}
N = \frac{1}{2} \left( 1 - S \right)
\end{equation}
with components only in the directions normal to the brane.

It is easy to check that the couplings (\ref{res}) are in agreement with those prescribed
by the action (\ref{SBIF}).

\subsection{Quadratic couplings}

There are two equivalent ways to reconstruct the quadratic couplings of the closed string
states with a D$p$-brane from the conformal theory. One approach is to use the boundary
state, sew it with a string propagator and saturate it with the vertices of the two
external states $S_1$ and $S_2$. This approach has been used in \cite{9808074} to obtain
the anomalous coupling of a D-brane with the $B_{\mu\nu}$ field and also non-anomalous
couplings with curvature terms \cite{9812149}. The other approach is instead to evaluate
the correlator of the vertices on the disk. This latter method has been discussed in
\cite{9603194,9611214,9809100} for ordinary D-branes of type II and extended in
\cite{9901085} for branes of type 0.

In the boundary states approach, the amplitude describing the coupling of two closed
string states is given by \footnote{ The projective invariance of the sphere has allowed
us to fix the position of the $S_2$ vertex at $z=\zb=1.$}:
\begin{eqnarray}
 A_{S1,S2} &=& C_0^{sp} \hat{N}^3 \
 \langle V_{S_1}~|
 \ V_{S_2}(1,1) \ \Delta
 \ |~{\rm D}p \rangle
\end{eqnarray}
where
\begin{eqnarray}
& C^{sp}_0 ~=~ \frac{4\pi}{\ap}\left( \frac{\pi}{\ks}\right)^2
 \hskip1cm , \hskip1cm
 \hat{N} ~=~ \frac{1}{\pi} ~ \ks
&
\end{eqnarray}
are the topological normalization respectively for the sphere and for the states
\cite{9601143}, and $\Delta$ is the usual closed string propagator:
\begin{eqnarray}
\label{eclprop}
 \Delta &=&
 \frac{\ap}{4\pi}
 \int_{|z|<1} \frac{d^2z}{|z|^2} z^{L_0} \bar{z}^{\tilde{L}_0}
\end{eqnarray}
The overlap equations defining $|{\rm D}p\rangle$ allow us to identify the right movers
with the left movers by means of the $S^{\mu}_{\ \nu}$ matrix (\ref{esse}) encoding the
information about the longitudinal and the transverse directions. After this
identification the right movers operators appear only in the boundary exponential and
annihilate the left or right vacua (for details see e.g. \cite{0004198}).

The other approach is to evaluate the correlator of the vertices on the disk
\cite{9603194,9611214}. Here one can use the {\sl doubling trick} to duplicate the region
inside the disk to fill the whole complex plane. This is done introducing the matrix $S$
(\ref{esse}). The conformal invariance of the disk allows us to fix three (real)
parameters for the four punctures and leaves just one modulus to be integrated over. In
this way we get the right correlator for the vertices, but we do not determine the right
normalization involving the tension of the brane. Perfect equivalence between the two
approaches is reached by adding in the latter case an overall multiplication factor $\ks~
N_{U,p}$ or $\ks~ N_{T,p}$ (see appendices \ref{aboundary} and \ref{avertices} for our
conventions) according to the required boundary sector (\ref{eboundary}).

\begin{figure}[t]
  \centering
  \epsfig{figure=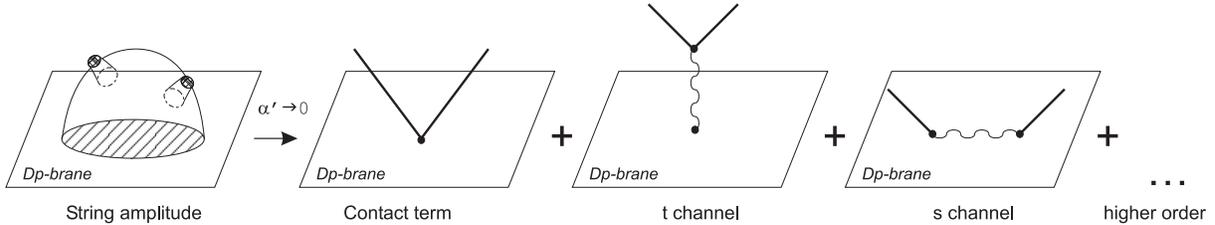,width=16cm}
\caption{The low energy expansion. The first term in the expansion represent the contact
term; this is directly the two point vertex present in the effective Lagrangian. However,
the two point string function contains also the other two exchange diagrams.}
  \label{fdisklowenergy}
\end{figure}

In order to get the interaction vertices of the boundary action, we must perform the low
energy limit $\ap \rightarrow 0$. It is crucial to realize that the string amplitudes in
this limit exhibit two types of divergences which are due to the exchange of massless
particles in the {\sl s-} and {\sl t-channel} (fig. \ref{fdisklowenergy}). So, we must
evaluate these diagrams and subtract them from the string amplitudes.

To evaluate the {\sl t-channel} exchange diagram (fig. \ref{fdisklowenergy}), we need the
$S_{IIA}$ bulk action (\ref{bulk}) suitably compactified \cite{0012035}. In fact we
evaluate it with the sewing technique: we start from the couplings of the boundary with
an intermediate state, propagate it in the bulk and finally sew with a three point
interaction vertex read from the compactified bulk action. Of course, the exact
cancellation of the divergences requires the sum on all possible intermediate states.

The {\sl s-channel} diagram represents the interaction of the external states with an
open string state living on the brane. In this case, the two interaction points can be
obtained by string theory techniques because now they are irreducible. So, for each
external states we correlate its vertex with an intermediate open string vertex, and then
sew together with the proper propagator.

In ten dimensions this technique has been successfully applied in \cite{9809100} to
reproduce the world volume action (\ref{BI}). We want to extend it to the orbifold
compactification in order to reproduce the action describing the world volume
interactions of the fractional branes (\ref{SBIF}). In this case there is a larger
variety of states and we summarize in table \ref{tqcoup} all possible quadratic
couplings. For shortness and notation convenience we introduce a $4\!\times \!4$ matrix
$Q$ according to table \ref{tqcoup}, to easily identify the sectors we will refer to. For
example the symbol $Q_{23}$ means a string amplitude between a closed string of the R-R
untwisted sector, a closed string state of the NS-NS twisted sector and the boundary
state; its explicit expression is:
\begin{eqnarray}
 Q_{23} &=& C_0^{sp} \hat{N}^3 \ \ks N_{T,p} ~
 \langle ~ V_{NST}^{(0,-1)} ~|~ V_{RU}^{(-1/2,-1/2)} (1,1)~
 \Delta
 ~|~ B_{RT}^{(-3/2,-1/2)} ~\rangle
\end{eqnarray}

\begin{table}[t]
\begin{equation*}
\begin{array}{||c|cc|cc||}
\hline\hline
		& NS,U     & R,U      & NS,T      & R,T     \\ \hline
  NS,U  & B_{NSU}  & B_{RU}   & B_{NST}   & B_{RT}  \\
  R,U   &          & B_{NSU}  & B_{RT}    & B_{NST} \\ \hline
  N,U   &          &          & B_{NSU}   & B_{RU}  \\
  R,T   &          &          &           & B_{NSU} \\ \hline\hline
\end{array}
\end{equation*}
\caption{Here we summarize all the possible quadratic couplings. In the columns there are
the possible sector for particle 1; the rows are for particle 2. For each couple of
states we report the only interacting sector of the D$p$-brane.} \label{tqcoup}
\end{table}

Each element of this matrix $Q$ must reproduce some of the terms of the boundary action.
So, for example, we have:
\begin{eqnarray}
 Q_{11} & \rightarrow &
	e^{-\frac{1-p}{2}\phi}\sqrt{-\ {\rm det} (\tilde{G}+ \esp^{-\phi}\tilde{B})} \Big|
	_{\text{\tiny{\begin{tabular}{c}quadratic\\ coupling\end{tabular}}}}
	\nod
 Q_{12} & \rightarrow & C_{p-1} \wedge B \nod
 Q_{22} & \rightarrow & 0 \no
\end{eqnarray}

The calculation of these terms and their correspondence with the appropriate boundary
action has been discussed respectively in \cite{9809100}, \cite{9808074} and
\cite{9603194} for the case of a flat background in ten dimensions. Here we have
performed the same calculation in the case of the orbifold compactification and, apart
from some obvious changes, the same structure is recovered. We will concentrate here on
the amplitudes involving one particle in a twisted sector, which are specific of the
fractional branes; we consider:
\begin{eqnarray}
 Q_{13} &=& C_0^{sp} \hat{N}^3 \ \ks N_{T,p} ~
 \langle ~ V_{NST}^{(-1,-1)} ~|~ V_{NSU}^{(0,0)} (1,1) ~
 \Delta
 ~|~ B_{NST}^{(-1,-1)} ~\rangle
 \nod
 Q_{14} &=& C_0^{sp} \hat{N}^3 \ \ks N_{T,p} ~
 \langle ~ V_{RT}^{(-1/2,-3/2)} ~|~ V_{NSU}^{(0,0)} (1,1)~
 \Delta
 ~|~ B_{RT}^{(-3/2,-1/2)} ~\rangle
 \\
 Q_{23} &=& C_0^{sp} \hat{N}^3 \ \ks N_{T,p} ~
 \langle ~ V_{NST}^{(0,-1)} ~|~ V_{RU}^{(-1/2,-1/2)} (1,1)~
 \Delta
 ~|~ B_{RT}^{(-3/2,-1/2)} ~\rangle
 \nod
 Q_{24} &=& C_0^{sp} \hat{N}^3 \ \ks N_{T,p} ~
 \langle ~ V_{RT}^{(-1/2,-1/2)} ~|~ V_{RU}^{(-1/2,-1/2)} (1,1)~
 \Delta
 ~|~ B_{NST}^{(-1,-1)} ~\rangle \no
\end{eqnarray}
The explicit expressions for the vertices are given in appendix \ref{avertices}.

For example, for the amplitude $Q_{13}$ we get:
\begin{eqnarray}\label{eq13}
Q_{13} &=& \ap ~ \ks N_{T,p} ~ \frac{\G(-\ap t/2)\G(2 \ap q^2)}
	{\G(1-\ap t/2+2 \ap q^2)} ~ \sqrt{2}
	\left(2q^2\,a_1+\frac{t}{2}\,a_2\right)
\end{eqnarray}
where
\begin{eqnarray}
\label{ecinema}
 t &=& -(p_1+p_2)^2 ~=~ -2p_1\inn p_2 \\
 q^2 &=& p_1\inn L\inn p_1 ~=~ {1\over2}p_1\inn S\inn p_1 \no
\end{eqnarray}
are respectively the momentum transferred to the brane and the longitudinal one.

The kinematic factors in (\ref{eq13}) are:
\begin{eqnarray}
	a_1&=& \beta \ ~ p_2\inn \pol_1 \inn p_2
\nod
	a_2&=& \beta \ \left[
		2(q^2-t/4) \, {\rm Tr}(\pol_1\inn S) \, +
		p_1 \inn S \inn \pol_1 \inn p_2 +
		p_2 \inn \pol_1 \inn S \inn p_1 +
		p_1 \inn S \inn \pol_1 \inn S \inn p_2
		\right] \no
\end{eqnarray}
where $\pol_1$ is the polarization of the first state in the NS-NS untwisted sector. So
by choosing the appropriate $\pol_1$ we select the graviton, Kalb-Ramond or dilaton field.

In the low energy limit $\ap \rightarrow 0$, we note the amplitude is divergent:
\begin{eqnarray} \label{eampdivergent}
Q_{13} &=& - ~ \ap ~ \ks N_{T,p} ~ \sqrt{2}
	\left(\frac{2}{t} \, a_1 + \frac{1}{2q^2} \, a_2 \right)
\end{eqnarray}

As explained before, we must subtract these divergences owing to the field theory
diagrams. Let's start with the t-channel. For definiteness we will consider the
interaction between $h_{\mu\nu}$ and $\beta$. The only three point interaction vertex we
can read from the bulk action is the kinetic term of the scalar $\beta$:
\begin{eqnarray}
 \frac{1}{2}\, (2 \, \ks \, h^{\mu\nu}) \,\partial_{\mu}\beta \, \partial_{\nu}\beta
\end{eqnarray}
Now we must sew the corresponding field theory amplitude $A^{bulk}=- 2 \ks \,
\pol_1^{\mu\nu} \, p^2_{\mu}\,\beta_2 \, p_{\nu}^3\, \beta_3 $ with the propagator $-
\frac{1}{t} $ for the twisted scalar and finally with the linear coupling with the brane
(\ref{elinbeta}):
\begin{eqnarray}
 A_t (h,\beta) &=&
 - 2 ~ p_2 \inn \pol_1 \inn p_2 \ \beta_2
 ~ \frac{1}{t} \
 \ks ~ \sqrt{2} N_{T,p}
\end{eqnarray}

Next, let's consider the s-channel. We need to evaluate the interaction between the
external states and an intermediate open string state $\lambda$ living and propagating on
the brane. These $\lambda^i$ fields represent the excitations of the transverse
coordinates $X^i$, whose expectation values correspond to the brane position; they appear
only in the pull back of the metric $\tilde{G}$. With our normalization for the boundary
action (\ref{SBIF}), from the kinetic term we can read the propagator:
\begin{eqnarray}
 \langle ~ \lambda^i \, \lambda ^j \, ~\rangle &=& - \ \frac{\sqrt{2} ~\ks}{\hat{T}_p} \ \frac{1}{p^2} \ N^{ij}
\end{eqnarray}

Now, instead of evaluating the interaction from the string, we will show an equivalent and
shorter way \cite{9809100}. The idea is to imagine the interaction of a generic state as
coming from the boundary action interpreted as the first order of a Taylor series in the
transverse coordinates. Considering the field $\beta$ as in the example, we can read the
coupling from the next to linear order of (\ref{elinbeta}):
\begin{eqnarray}
 V_{\beta, \lambda} &=& \sqrt{2} ~ N_{T,p} ~  V_{p+1} \ \lambda^j \partial_j \beta
\end{eqnarray}
Analogously, the interaction between $h$ and $\lambda$ gets two contributes:
\begin{eqnarray}
 V_{h, \lambda} &=& \frac{\hat{T}_p}{\sqrt{2}} ~
	\left[ \, {\rm Tr}(\pol_1 \inn L) \ p_1 \inn N \inn \lambda \ - \
	2 \ p_1 \inn L \inn \pol_1 \inn N \inn \lambda \,
	\right] \ .
\end{eqnarray}
The first is the next to linear order of the Taylor expansion and the second appears
directly in the pull back $2\, h_{ia} \partial^a X^i$. Sewing all together we get:
\begin{eqnarray}
 A_s (h,\beta) &=& - ~ \ks N_{T,p} \sqrt{2} \
 \frac{1}{q^2} ~ \beta ~
 \left[ \, p_2 \inn N \inn p_1 \ {\rm Tr}(\pol_1 \inn L) \ - \,
 2 \ p_2 \inn N \inn \pol_1 \inn L \inn p_1 \, \right]
\end{eqnarray}

Using the relation (\ref{long},\ref{enne}) and (\ref{ecinema}), we can subtract the
divergent part of (\ref{eampdivergent}) in all the amplitudes. In terms of the {\sl
standard} fields (\ref{edefcanonic}) directly appearing in the action (\ref{SBIF}), we
obtain:
\begin{eqnarray}
 &&
 Q_{13}(\phi,\tilde{b}) -
  A_t(\phi,\tilde{b}) -
  A_s(\phi,\tilde{b}) \ =\
	~ -\frac{N_{T,p}}{\ks} ~
	\frac{p-1}{2} \phi \ \tilde{b} \nod &&
 Q_{13}(h,\tilde{b}) -
  A_t   (h,\tilde{b}) -
  A_s   (h,\tilde{b}) \ =\
	-\frac{N_{T,p}}{\ks} ~
	{\rm Tr} \, ( \ks ~ h \inn L ) \ \tilde{b} \nod &&
 Q_{13}(B,\tilde{b}) -
  A_t(B,\tilde{b}) -
  A_s(B,\tilde{b}) \ =\
	0 \nod &&
 Q_{13}(\hat{\eta}_a,\tilde{b}) -
  A_t(\hat{\eta}_a,\tilde{b}) -
  A_s(\hat{\eta}_a,\tilde{b}) \ =\
	-\frac{N_{T,p}}{\ks} ~
	\frac{1}{2} \hat{\eta}_a ~ \tilde{b}
 \no
\end{eqnarray}

Similar calculations for the other massless closed string modes yield:
\begin{eqnarray}
 Q_{14} &=& \frac{N_{T,p}}{\ks} ~ \frac{1}{2~(p-1)!} ~
	A_{\mu_1\ldots\mu_{p-1}} ~ B_{\mu_p\mu_{p+1}} ~ \epsilon^{\mu_1\ldots\mu_{p+1}} \\
 Q_{23} &=& \frac{N_{T,p}}{\ks} ~ \frac{1}{(p+1)!} ~
	\tilde{b} ~ C_{\mu_1\ldots\mu_{p+1}} ~ \epsilon^{\mu_1\ldots\mu_{p+1}}
\end{eqnarray}

These are exactly the couplings we expect from the action (\ref{SBIF}).


\section{Conclusions}

By directly examining various string scattering amplitudes, we have extracted various
world-volume interactions for the effective field theory on fractional D-branes. Our
results are in perfect agreement with the Born-Infeld and Wess-Zumino terms of the action
(\ref{SBIF}), numerical coefficients included. A further check could be to evaluate the
quadratic coupling involving two twisted states explicitly.

As already emphasized, the techniques we have used are rather general. In particular it
could be interesting to apply them to the case of fractional branes extended also in the
compact directions. There are indeed many reasons to argue that the resulting space time
configurations are extended objects with a smaller density of mass and charges.


\section*{Acknowledgement}

We would like to thank M. Bill\'o, M. Frau, R. Russo, and in particular A. Lerda for very
useful discussions. G.S. thanks the {\sl Dipartimento di Fisica Teorica} of the {\sl
University of Torino} for the kind hospitality.

\appendix

\section{Notation \label{note}}

In the orbifold background (\ref{bg}) the Lorentz group $SO(1,9)$ is naturally broken to
$SO(1,5) \otimes SO(4)$. We reserve the Latin capital letters for $SO(1,9)$ indices, the
Greek ones for the six-dimensional spacetime and the Latin ones for the compact space.\\
We have then, for the ten dimensional vector representation:
\begin{eqnarray}
 X^{M} & \rightarrow & \{ X^{\mu},X^{i} \} \hskip1cm
	\left(
	\begin{array}{ccc}
	M   &= & 0 \ldots 9 \\
	\mu &= & 0\ldots 5 \\
	i   &= & 6 \ldots 9
	\end{array} \right)
\end{eqnarray}
and for the spinorial representation \footnote{ We recall the orbifold projection selects
just one chirality for $SO(4)$.} :
\begin{eqnarray}
	\begin{array}{ccc}
	S^A         & \rightarrow & \{S^{\alpha}\otimes S^a \}      \\
	S^{\dot{A}} & \rightarrow & \{S^{\dot{\alpha}}\otimes S^a \}
	\end{array}
  \hskip1cm
	\left(
	\begin{array}{ccc}
	A,\dot{A} &= &1 \ldots 32 \\
	\alpha,\dot{\alpha} &= & 1 \ldots 8 \\
	a &= & 1 \ldots 4
	\end{array}
	\right)
\end{eqnarray}

\section{Boundary normalization \label{aboundary}}

In each sector of the theory we can construct the boundary state:
\begin{eqnarray}
 |B,k,\eta\rangle & = &
  \esp^{i\theta} ~{\rm exp}
	\Bigg(\sum_{l>0}\Big[\frac{1}{l}\alpha_{-l}^{\mu}S_{\mu\nu}\tilde{\alpha}^{\nu}_{\-l}
					\Big]
		~ + ~ i\eta \sum_{m>0}\Big[\psi_{-m}^{\mu}S_{\mu\nu}\tilde{\psi}^{\nu}_{\-m}
					\Big]
	\Bigg)
\nod
	&&
	|B,k,\eta\rangle^{(0)}
\end{eqnarray}
where $l$ and $m$ are integer or half-integer depending on the sector, $k$ denotes the
momentum of the ground state and $\theta$ is a phase equal to $\pi$ in the untwisted R-R
and twisted NS-NS sectors, to $3\pi/2$ in the untwisted NS-NS sector and to $0$ in the
twisted R-R sectors. The parameter $\eta=\pm 1$ describes the two different spin
structures \cite{Pol,Cal}, and the matrix $S$ (\ref{esse}) encodes the boundary conditions
of
the D$p$-brane which we shall always take to be diagonal.\\
 We omit always the ghost and
superghost part of the boundary states, which are as in the usual case
\cite{9707068,9802088}. In order to obtain a localized D$p$-brane (say in $y=0$), we have
to take the Fourier transform of the above boundary state, where we integrate over the
directions transverse to the brane. For the untwisted sector we obtain:
\begin{eqnarray}
 |B,y=0,\eta\rangle &=& \prod_{\nu=6}^{9}\Big( \sum_{n_{\nu}} {\rm
 e}^{i\frac{n_{\nu}}{R_{\nu}}\hat{q}}\Big)~\frac{1}{(2\pi)^{5-p}} ~ \int d k^{5-p} ~ {\rm
 e}^{ik\hat{q}} ~|B,k,\eta\rangle,
\end{eqnarray}
while for the twisted one:
\begin{eqnarray}
 |B,y=0,\eta\rangle &=& \frac{1}{(2\pi)^{5-p}} ~ \int d k^{5-p} ~ {\rm
 e}^{ik\hat{q}} ~|B,k,\eta\rangle.
\end{eqnarray}

The invariance of the boundary state under the GSO and the orbifold projection always
requires that the physical boundary state is a linear combination of the two states
corresponding to $\eta=\pm$. In the conventions of \cite{9707068,9802088} these linear combinations are of
the form:
\begin{eqnarray}
 |B\rangle_{NS,U}   &=& \frac{1}{2}\Big( |B,+\rangle_{NS,U} ~-~|B,-\rangle_{NS,U}\Big)\\
 |B\rangle_{R,U}    &=& \frac{1}{2}\Big( |B,+\rangle_{R,U}  ~+~|B,-\rangle_{R,U}  \Big)\\
 |B\rangle_{NS,T}   &=& \frac{1}{2}\Big( |B,+\rangle_{NS,T} ~+~|B,-\rangle_{NS,T}\Big)\\
 |B\rangle_{R,T}    &=& \frac{1}{2}\Big( |B,+\rangle_{R,T}  ~+~|B,-\rangle_{R,T}  \Big)\\
\end{eqnarray}

A fractional D$p$ brane state can be written as:
\begin{eqnarray}
\label{ebound}
 |{\rm D}p\rangle &=& N_{U,p} ~
	\Big( |B\rangle_{NS,U} \pm |B\rangle_{R,U} \Big)
	\pm N_{T,p} ~
	\Big( |B\rangle_{NS,T} \pm |B\rangle_{R,T} \Big)
\end{eqnarray}

Now we want to determine the constant $N_{T,U}$ solving the open-closed consistency
condition. We have to compare the closed string cylinder diagram with the open string
one-loop diagram. We obtain:
\begin{eqnarray}
 N_{U,p} &=& \frac{1}{2\sqrt{2}}\hat{T}_p\\
N_{T,p} &=& \sqrt{2\pi} (4\pi^2 \ap)^{\frac{1-p}{2}}
\end{eqnarray}
where
\begin{eqnarray}
 \hat{T}_p &=& \frac{T_p}{\sqrt{V_4}} \ \ \ \ \ {\rm and} \ \ \ \ \ T_p=\sqrt{\pi}(4\pi^2
 \ap)^{\frac{3-p}{2}}
\end{eqnarray}

\section{Vertices \label{avertices}}

According to the notation established in appendix (\ref{note}), we write the following
vertices:

\begin{eqnarray}
 && \mbox{\bf NS-NS Untwisted} \nod
	V_{NSU}^{(0,0)} (z,\zb) &=&
		\varepsilon_{MN} \
		[
		\partial X^{M}+i (k\cdot \psi)\psi^{M}
		]_{z}
		[
		\partial \tilde{X}^{N}+i (k\cdot \tilde{\psi})\tilde{\psi}^{N}
		]_{\zb}
		\ U^{(0,0)} (z,\zb) \\
	\mbox{where} &&
		\begin{array}{lll}
		h~:\ &
			\varepsilon^h_{\mu\nu} = +~ \varepsilon^h_{\nu\mu} &
			\varepsilon^h_{ij} =~0 \\
		B~:\ &
			\varepsilon^B_{\mu\nu} = -~ \varepsilon^B_{\nu\mu} &
			\varepsilon^B_{ij} =~0 \\
		\varphi~:\ &
			\varepsilon^{\varphi}_{\mu\nu} = \frac{1}{\sqrt{6-2}}
			\left(\eta_{\mu\nu} - p_{\mu}l_{\nu}-l_{\mu}p_{\nu} \right) &
			\varepsilon^h_{ij} =~0 \\
		\eta_k~:\ &
			\varepsilon^{\eta_k}_{\mu\nu} = ~0 &
			\varepsilon^{\eta_k}_{ij} = \delta^k_i \delta^k_j \\
		\end{array}
\nod
\nod
 && \mbox{\bf NS-NS Twisted} \nod
	V_{\beta}^{(-1,-1)} (z,\zb) &=& \frac{1}{\sqrt{2}}
		C_{ab}S^a(z)\bar{S}^b(\zb)
		\ T^{(-1,-1)} (z,\zb)
\\
\nod
 && \mbox{\bf R-R Untwisted} \nod
	V_{RU}^{(-1/2,-1/2)}(z,\zb) &=&
		F_{A\dot{B}}
		S^A(z)\bar{S}^{\dot{B}}(\zb)
		\ U^{(-\frac{1}{2},-\frac{1}{2})} (z,\zb) \\
	\mbox{where} &&
		F_{A\dot{B}} = \sum_{n~even} ~ \frac{1}{n!} ~ (C\Gamma^{\mu_1\cdots \mu_n})_{A\dot{B}} ~ F_{\mu_1\cdots \mu_n}
\nod
\nod
 && \mbox{\bf R-R Twisted} \nod
	V_{RT}^{(-1/2,-1/2)}(z,\zb) &=&
		G_{A\dot{B}}
		S^{\alpha}(z)\bar{S}^{\dot{\beta}}(\zb)
		\ T^{(-\frac{1}{2},-\frac{1}{2})} (z,\zb)
		\\
	\mbox{where} &&
		G_{\alpha\dot{\beta}} = \sum_{n~even} ~ \frac{1}{n!} ~ (C\Gamma^{\mu_1\cdots \mu_n})_{\alpha\dot{\beta}}
			~ G_{\mu_1\cdots \mu_n}
\nod
\nod
 && \mbox{and} \nod
	U^{p,q}(z,\zb) &=& \esp^{p\phi} \esp^{q\tilde{\phi}} \ \esp^{i k\cdot X(z,\zb)}
	\nod
	T^{p,q}(z,\zb) &=& \esp^{p\phi} \esp^{q\tilde{\phi}} \ \Lambda(z)~\tilde{\Lambda}(\zb) \ \esp^{i k\cdot X(z,\zb)}
	\no
\end{eqnarray}


\end{document}